\documentclass{ws-p8-50x6-00}

\begin{document}

\title{Chemical evolution of the Galaxy: The G-dwarf problem and radioactive 
chronology revisited taking account of the Thick Disk}   

\author{B.E.J. Pagel}  

\address{Astronomy Centre, CPES, University of Sussex, Brighton BN1 9QJ, UK 
\\E-mail: bejp@star.cpes.susx.ac.uk}


\maketitle

\abstracts{
Thin and Thick disk metallicities overlap, but 
$\alpha$(Fe) relations differ;  there is  also a hiatus 
in time before the first thin-disk stars, with a minimum metallicity 
[Fe/H] $\simeq -0.6$.   Thus there are two `G-dwarf' problems. We fit  
metallicity distribution functions (MDFs) for the two disks derived by Wyse \& 
Gilmore (1995) with ana\-lytical models (for O and Mg).  The Thick disk fits a 
simple inflow model with  effective yield and final [Mg/H] about the same as 
for the Thin disk, excluding continuous mass loss from the former as a source 
of `prompt initial enrichment' (PIE) for the latter. The Thin disk has a 
narrow MDF fitted (rather poorly) 
with a closed model having PIE from some other source, or more elegantly 
with a slow infall model similar to the `two-inflow' scenario of 
Matteucci et al., but with a PIE and a severe break in the age-metallicity 
relation. Implications for radio-active cosmochronology  are investigated.  
}

\vspace{-7mm} 
\section{Introduction}

The `G-dwarf' problem 
has been with us for nearly 40 years 
and extended to other galaxies.  
There is still no definitive universally accepted solution; rather too 
many, the most popular being `prompt initial enrichment' (PIE; Truran \& 
Cameron 1971) and inflow (Larson 1972).  Truran \& Cameron postulated a 
first generation of massive stars, sometimes called Population III, 
collapsing to black holes or `collapsars' in the halo, thereby accounting for 
its dark mass. 
Later the idea gained ground that the 
halo consists mostly of dark matter, wholly or partly non-baryonic, but 
PIE came in by another route.  After discovery of the Thick Disk,  
Gilmore \& Wyse (1986) 
suggested that (a) the Thick disk was initially enriched by 
gas shed from the halo as in the model by Hartwick (1976); and (b) 
the Thin disk was 
similarly enriched by gas shed from the Thick one. I never took much notice of 
this paper, because the relative numbers are not right unless there is 
inflow as well, and Wyse \& Gilmore (1992) themselves abandoned proposition 
(a) on angular momentum grounds. This still leaves open   
a connection between the two disks, and Wyse \& Gilmore (1995) have published 
separate MDFs for them which, with new data on O/Fe and Mg/Fe, have 
implications that I discuss in this talk.       

\vspace{-3mm} 
\section{Iron and magnesium MDFs and AMRs}

\begin{figure*}[htbp] 
\vspace{3.5cm} 
\includegraphics{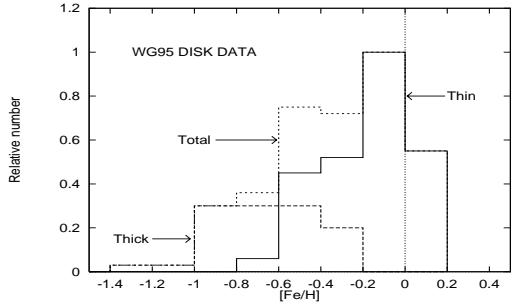} 
\caption{Iron MDFs for Thick and Thin disks, after Gilmore \& Wyse (1995).}  
\end{figure*}

Numerous models based on inflow give a fair fit to the 
MDF of nearby disk stars as a whole (Matteucci \& Fran\c cois 1989; 
Pagel 1989; Pagel \& Tautvai\v sien\. e 1995).  
PIE models based on outflow of metals 
from a rapidly evolving bulge have also been quite successful (K\" oppen 
\& Arimoto 1990; Samland et al.\ 1997). Strobel (1991), Pardi et al.\  
(1995), Chiappini et al.\ (1997),  
Fuhrmann (1998) and Gratton et al.\ (2000), among others, 
have investigated chemical evolution implications of the different 
time-scales of evolution of the two disks, and Matteucci and co-authors 
have developed `two-inflow' models which I consider a good idea, but 
I shall argue that they do not go far enough.  

\begin{figure*}[htbp] 
\vspace{4.5cm}
\includegraphics{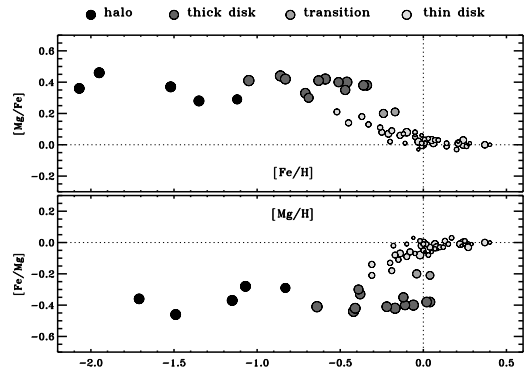} 
\caption{[Mg/Fe] vs [Fe/H] and [Fe/Mg] vs [Mg/H] for stars of the Galactic halo,
thick disk and thin disk, after Fuhrmann (1998). Courtesy Klaus Fuhrmann.}
\end{figure*} 

 Wyse \& Gilmore's iron MDFs are shown in Figure 1, where it is seen   
that the Thick disk has a mean [Fe/H] just at the sharp lower limit of 
the Thin one, which made it seem natural for Gilmore \& Wyse (1986) 
to suggest a Hartwick-like model for the Thick disk where gas expelled 
from it over time led to a low effective yield and had just the right 
metallicity to start off the Thin disk, the mass of which can be made  
up as desired by subsequent inflow.  This, however, assumes a single 
$\alpha$(Fe) relation, which now seems incorrect (Fuhrmann 1998; 
Gratton et al.\  
2000), as is illustrated in Figure 2. The Thick-disk stars have enhanced 
$\alpha$/Fe (and O/Fe) even when their $\alpha$-metallicity is as high as solar.  

\begin{figure*}[htbp]
\vspace{4.5cm}
\includegraphics{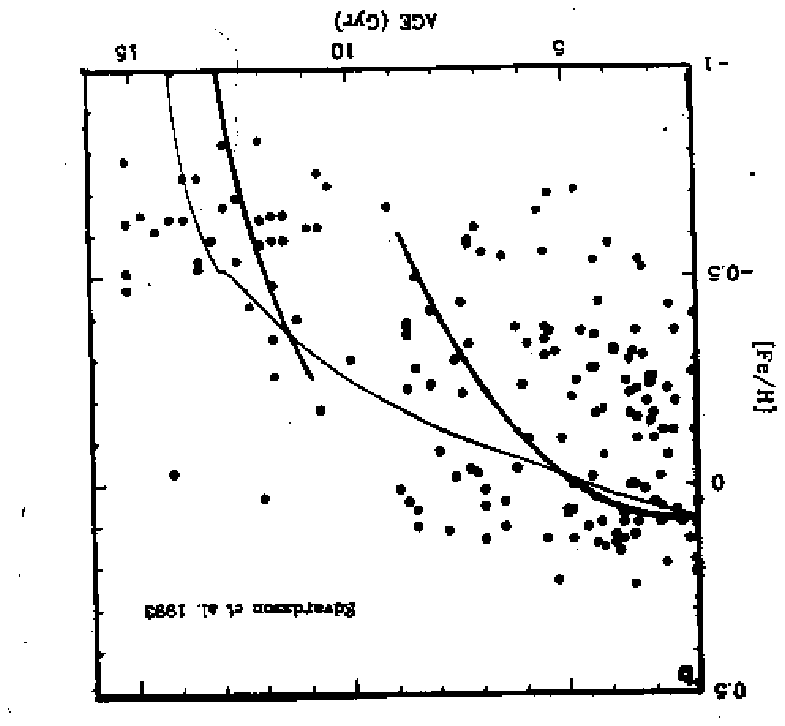}
\caption{Thin curves: age-metallicity relation from the `two-inflow' model of 
Chiappini, Matteucci \& Gratton (1997).  Thick curves: sketch of suggested 
separate AMRs for the Thick and Thin disks. Adapted from Chiappini et al.\ 
(1997).}   
\end{figure*}

This gives further support to a short time-scale for Thick-disk star formation, 
confirmed by HIPPARCOS 
parallaxes (Ng \& Bertelli 1998), which show a hiatus between that and the 
first stars of the Thin disk, also visible in the data from Edvardsson et al.\ 
(1993) shown in Fig 3. The need for two time-scales and a hiatus in star 
formation, as postulated by 
Chiappini et al., is clearly evident, but their model does  
not go far enough, and it seems more realistic to draw 
in two entirely separate age-metallicity relations as shown schematically 
by the thick curves. 

We also face the complication that the iron-to-$\alpha$-element conversion 
is different in the two disks.  Making use of the relationships 
embodied in Fig 2, we can convert Fig 1 into the Mg MDF shown in Fig 4.

\begin{figure*}[htbp]
\vspace{4.5cm}
\includegraphics{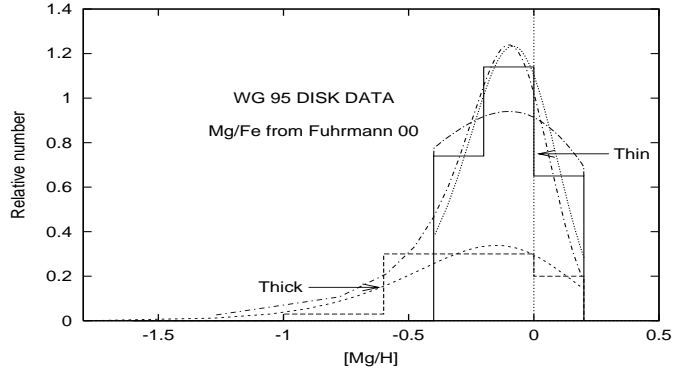}
\caption{Magnesium MDFs. Histograms: data after Wyse \& Gilmore (1995). 
Curves: dashed, simple inflow model for Thick disk; dotted, PIE plus 
inflow model for Thin disk; dash-dotted, inflow model without PIE; 
long-dash dotted, PIE without inflow.}  
\end{figure*}

Fig 4 shows some interesting features not previously apparent.  For one thing, 
the upper abundance limits are identical and the means nearly so, so there is 
no longer any basis for assuming Hartwick-style mass outflow 
from the Thick disk with a resulting reduction in effective yield.  
It does have  a low-metallicity tail, also 
noticed by Beers \& Sommer-Larsen (1995), whereas the Thin one has a sharp lower 
limit suggestive of PIE; but if the previous metallicity came from the Thick 
disk I would rather call it Prompt Initial Impoverishment!  

The curves in Fig 4 show some analytical fits to the two MDFs.  The Thick disk 
is readily fitted by a simple decaying inflow model as in Pagel (1997) with a 
yield 0.15 dex below solar, a final gas fraction of 0.1 and any time scale less 
than that of SNIa, say around 1 Gyr.   

For the Thin disk there are more choices. A pure PIE model without inflow gives 
a barely acceptable fit, while a pure inflow model (Clayton 1985 standard model 
with $k=4$) produces too much of a low-metallicity tail. A combined PIE-inflow 
model fits very well, possibly representing terminal Thick-disk gas diluted by 
inflow of unprocessed material before the Clayton episode --- thus a 
`three-inflow' model.  

\vspace{-3mm} 
\section{Implications for radio-active cosmochronology} 

Solar-system actinide abundances have been used in the past to guess the age of 
the Galaxy, although there is also a clear dependence on chemical evolution models. 
More recently, with the discovery of r-process rich halo stars, it has become 
possible to use stellar thorium abundances in a manner that is less model-dependent,
but the age of the halo does depend explicitly or implicitly on the Th/Eu production 
ratio. This has been estimated on the basis of nuclear physics, but with a fair 
degree of uncertainty (see Westin et al.\ 2000), while Truran (1999) has appealed 
to the proto-solar ratio, but this is clearly only a lower limit, thereby giving 
just a minimum possible age for the halo stars. 

Now we know more about stellar ages and the gap between the two disks, 
it may be of interest to turn the argument round and check for consistency. We use 
our Galactic models to estimate the $K$-ratios defined by Fowler (1987): 
{\small 
\begin{equation}
K_{ij} \equiv \frac{N_ i}{N_ j}(\rm {protosolar})/\frac{p_ i}{p_ j} 
\end{equation}
} 
and use the resulting $K_{{\rm Eu\;Th}}$ to predict the production ratio from the 
known proto-solar abundance ratio. 

We recall Fowler's own model, a closed PIE model from which he deduced an age for 
the disk of just 10 Gyr.  If we think only of the Thin disk, then this was 
remarkably prescient, but now we also have to think of the Thick disk or bulge 
where any PIE would probably have originated and which is 2 or 3 Gyr older. 
We assume the Thick disk to have formed instantaneously 12 Gyr ago and the Thin 
disk to have started forming after an interval $\epsilon = 3$ Gyr. The Solar 
System was formed after a further interval $\Delta = 4.4$ Gyr.  Fowler's equations 
are then modified to:
{\small 
\begin{eqnarray} 
N_ i ({\rm radioactive})& \propto & p_ i\left[\frac{S}{1-S}e^{-\lambda_ i(\Delta 
+\epsilon)} + \frac{1-e^{-\lambda_ i\Delta}}{\lambda_ i\Delta}\right] \\
N_ k({\rm stable})&\propto& p_ k/(1-S) 
\end{eqnarray}
}  
where in our case the PIE contribution $S=0.4$ rather than 0.17. Corresponding 
formulae for the 
inflow model have been given by Clayton (1985). Resulting $K$-ratios in the 
Solar System  are given in Table 1, where it is seen that the modified Fowler 
model still gives excellent agreement with `observed' $K$-ratios, whereas the 
inflow model gives numbers that are barely compatible if at all and presumably 
represents a generous upper limit to the modifications from inflow.         

\begin{table}[t]
\small  
\caption{$K$-ratios of actinide abundances at Solar System formation}   
\begin{center}
\begin{tabular}{|r|l|l|c|}
\hline
& 235, 238 & 232, 238 & Eu, Th \\ \hline  
`Observed' after Fowler (1987) & $0.25\pm .04$ & $1.35\pm .10$ & \\ 
Modified Fowler model & 0.24 & 1.45 & 1.23\\ 
Inflow model (Clayton $k=4$) & 0.49 & 1.13 & 1.06\\    
\hline
\hline
\end{tabular}
\end{center}
\end{table}

Implications for halo ages are given in Table 2.  Fowler's model implies a large 
correction from protosolar to production ratio and hence a large (and perhaps 
unacceptable) figure for the halo age, while the inflow model gives only a 
minimal correction and a more acceptable age range.  It is intriguing that the 
range of production ratios permitted by Galactic chemical evolution models 
matches so closely the range deduced from nuclear physics.   

\vspace*{-5mm} 
\begin{table}[t] 
\small   
\caption{ Th/Eu ratios and halo age}  
\begin{center}
\begin{tabular}{|r|c|c|}
\hline
& Th/Eu ratio & halo age (Gyr) \\ \hline  
Halo stars & 0.22 to 0.25 &\\
Proto-solar & 0.46 &\\ 
Nuclear theory prediction & 0.48 to 0.55 & 13.2 to 18.6\\ 
Modified Fowler model prediction & 0.57 & 16.7 to 19.3\\
Inflow model prediction & 0.49 & 13.7 to 16.3\\
\hline
\hline
\end{tabular}
\end{center}
\end{table}

\end{document}